\documentclass[twocolumn,showpacs,amsmath,amssymb,superscriptaddress,email]{revtex4-2}

\usepackage{graphicx}
\usepackage{dcolumn}
\usepackage{bm}

\usepackage[utf8]{inputenc}
\usepackage[T1]{fontenc}
\usepackage{mathptmx}
\usepackage{upgreek}
\usepackage{xcolor}
\usepackage{hyperref}
\hypersetup{
	colorlinks = true,
	linkcolor  = blue,
	filecolor  = blue,
	citecolor  = blue,
	urlcolor   = blue,
}
\begin{document}

\preprint{AIP/123-QED}

\title{Float, borosilicate and tellurites as cover glasses in Si photovoltaics: optical properties and performances under sunlight}

\author{M. P. Belançon}
    \email{marcosbelancon@utfpr.edu.br}
\author{M. Sandrini}
\author{H. S. Muniz}
\affiliation{Departamento de Física, Universidade Tecnol\'ogica Federal do Paran\'a, Pato Branco, PR  85503-390, Brazil}
\author{L. S. Herculano}
\author{G. V. B. Lukasievicz}
\affiliation{Departamento de Física, Universidade Tecnol\'ogica Federal do Paran\'a, Medianeira, PR 85884-000, Brazil}

\author{E. L. Savi}
\author{O. A. Capeloto}
\author{L. C. Malacarne}
\author{N. G. C. Astrath}
\author{M. L. Baesso}
\affiliation{Departamento de Física, Universidade Estadual de Maringá, Maring\'a, PR 87020-900, Brazil}

\author{G. J. Schiavon}
\author{A. A. Silva Junior}
\affiliation{Mestrado Profissional em Inovações Tecnológicas, Universidade Tecnol\'ogica Federal do Paran\'a, Campo Mourão, PR  87301-899, Brazil}

\author{J. D. Marconi}
\affiliation{%
Centro de Engenharia, Modelagem e Ciências Sociais Aplicadas, Universidade Federal do ABC, Santo André, SP 09210-580, Brazil
}%

\date{\today}

\begin{abstract}
One of the most significant materials in a solar panel is the glass, which provides transparency, UV protection as well as mechanical and chemical resistance. In this work, we describe the production of prototypes of four solar modules made using borosilicate, zinc-tellurite, Pr$^{3+}$ doped zinc-tellurite, and float glass as cover materials. The performance of these prototypes was evaluated under a solar simulator, and a device was developed to monitor all prototypes under real conditions. A comparison between indoor and outdoor measurements shows that outdoor results are fundamental to evaluate the performance of modified solar modules as the ones considered in this study. In addition, we demonstrate the fundamental role played by the refractive index of cover glasses in the performance of the prototypes, and discuss how this feature could be explored to achieve enhanced devices, as well as other benefits that may arise from this field of research.\end{abstract}

\maketitle

In the 1950s two breakthroughs made today's mainstream solar energy technology possible: Silicon solar cells~\cite{Pearson1957} and the float glass method to produce flat glasses~\cite{Pilkington1969}. Before that time, high-quality flat glass was produced by the metal plate manufacturing process. Due to the need of grinding and polishing, this method wasted about 20\% of the glass mass. It demanded high capital and operating costs, which motivated the development of the float glass process, where a ribbon of glass floats on the surface of a molten tin bath while continuously moving and cooling. This process produces highly homogeneous flat glasses, making grinding and polishing unnecessary. The technology was ready to dominate the market at the end of that decade, after about 7 years under development at the Pilkington industries~\cite{Pilkington1969}. A modern floating line produces flat glass at speeds of $\sim$4000 m$^2$/h and coatings can be cheaply deposited inline by chemical vapor deposition (CVD) at fast rates~\cite{McCurdy1999} ($\sim$100 nm/s), providing for example anti-reflective or self-cleaning capabilities~\cite{Sarkn2020}. Silicon photovoltaics are profiting from float glass transparency, mechanical quality, chemical resistance, and UV protection, which are fundamental to produce panels that achieve a lifespan of several decades. 

Silicon photovoltaics (PV) constituted 95\% of the photovoltaic market in 2020, which corresponds to about 200 GW$_p$ in annual production capacity~\cite{ITRPV2019}. The power per area in such devices is $\sim 200$ W/m$^2$, and roughly 1 billion m$^2$ of solar panels are produced every year. Even though it seems impressive, our needs are orders of magnitude beyond this number, which imposes many challenges towards the tera-watt deployment of PV~\cite{Haegel2019}. One of them is the expansion of production of float glass~\cite{Burrows2015}. The thickness of the cover glass in most of the PV on the market today is above 3 mm, and bi-facial PV~\cite{Kopecek2018} growth is expected to increase~\cite{ITRPV2019} and speed up the flat glass demand. Pilkington's method was decisive to enable this massive deployment of solar energy, and glasses may contribute even further to tackle the climate crisis.

About 80 million tonnes of solar panels are going to reach their end-of-life (EOL) in the next 30 years~\cite{Heath2020}. Glasses account for more than 50\% of this total weight~\cite{Xu2018}, and glass science may contribute to reducing this amount of EOL products. There is a demand to reduce the thickness~\cite{Allsopp2020} (and weight) of cover glasses in PV's, as well as to attenuate their degradation~\cite{Katayama2019}, which consequently would expand PV's lifespan even further. This would also reduce the demand for metals, which has been pointed out as a constraint to PV expansion and sustainability~\cite{Graedel2015a, Graedel2015, Manberger2018a, Li2019, Ren2021}. Although the cover glass may seem one of the simplest components of a PV panel, several complex features are demanded of a material to be viable in practical applications. Besides high transparency, chemical, mechanical, and UV resistance~\cite{Allsopp2020}, other properties include coatability~\cite{Deubener2009}, which is fundamental to reduce reflection losses in the glass/air interface and may even be explored to provide self-cleaning capabilities to the surface~\cite{Jesus2015}.

During the last decade, a significant effort has been made to develop spectral converters (SC), which could be built-in the cover glass or deposited as a thin film over or below it, to improve the spectral mismatch between sunlight radiation and Si solar cells spectral sensitivity. SC's provide by this way a mechanism to expand the efficiency limit of PVs beyond the Shockley-Queisser limit~\cite{Markvart2019}. This can be achieved by processes such as downconversion (DC)~\cite{Liu2009a,Leonard2013,DelaMora2017a} or upconversion~\cite{Goldschmidt2015,Fischer2018}. The downshifting (DF) of photons, a phenomenon that can be classified as a subcategory of DC~\cite{DelaMora2017a}, consists in the conversion of high energy photons into low energy photons, with a quantum efficiency lower than 1. Pr$^{3+}$ doped materials, for instance, offer one of the simplest ways to achieve DF, due to the well-known emission lines of this ion around ~600 nm, ~650 nm and ~1050 nm under blue light excitation \cite{Belancon2014c}. Yet interesting scientific findings have been reported on several SC materials, most of them are unfeasible at the needed scale~\cite{Huang2013,McKenna2017a,DelaMora2017a,Bubli2020a,Ferreira2020}. In this way, some authors have investigated small variations in the composition of the float glass~\cite{Allsopp2020}, which should not interfere in the production process and would be readily available at the industry level.

Several works have demonstrated high quantum efficiencies of tellurite glasses, a glass family that we have been investigating in the last few years~\cite{Belancon2014c,Taniguchi2019, Taniguchi2020a}, and could potentially be used as SCs~\cite{Han2015a, Zhou2016}. A few prototypes made with glasses of this family were already investigated~\cite{Garcia2019a} and, to the best of our knowledge, they were not evaluated under sunlight or compared to other prototypes made with commercially available glasses, which is the goal of this work. Here we report on the evaluation of tellurite glasses as cover materials of PV prototypes, where they are directly compared to similar prototypes made with two commercially available glasses. To accomplish this comparative study, we also developed an affordable tool to monitor the power output of several prototypes simultaneously, which was used to evaluate their performances under real sunlight. The following glasses were selected for this investigation: low iron float glass; borofloat 33; the TZNL (73.3\%Te-19.6\%Zn-4.9\%Na-2.2\%La~\cite{Taniguchi2020a}) and Praseodymium doped (0.1\%Pr$_{6}$O$_{11}$) TZNL glasses. Some key properties of these samples are shown in Table~\ref{tab:1}.  

\begin{table*}
\caption{\label{tab:1} Properties of the glasses investigated in this work. TZNL data that includes an error estimate were obtained by the thermal mirror method as described in Ref.~\onlinecite{Zanuto2013}.}
\begin{ruledtabular}
\begin{tabular}{lccccccc}
Glass&$\rho$ (g/cm$^3$)&$n$\footnote{Linear refractive index at 633 nm}&Thickness (mm)&k (Wm$^{-1}$K$^{-1}$)&$c$ (J kg$^{-1}$K$^{-1}$)& D ($10^{-7}$m$^2$s$^{-1}$)&$\alpha (10^{-6}$K$^{-1})$\\
\hline
Float glass\footnote{Data from Saint-Gobain} & 2.5 & 1.52 &4.0&1.06&800&5.3&9\\
Borosilicate 3.3\footnote{Data from Schott} & 2.23 & 1.47&1.54&1.2&830&6.8&3.3\\
TZNL& 5.34$\pm0.01$ & 1.98 (Ref.~\onlinecite{Serdang2009})&4.2&1.1 (Ref.~\onlinecite{Kutin2019})&801$\pm85$&2.6$\pm0.2$&11.8$\pm0.8$\\
\end{tabular}
\end{ruledtabular}
\end{table*}

\begin{figure}
\includegraphics[trim=1cm 0.5cm 1cm 1cm,scale=0.5]{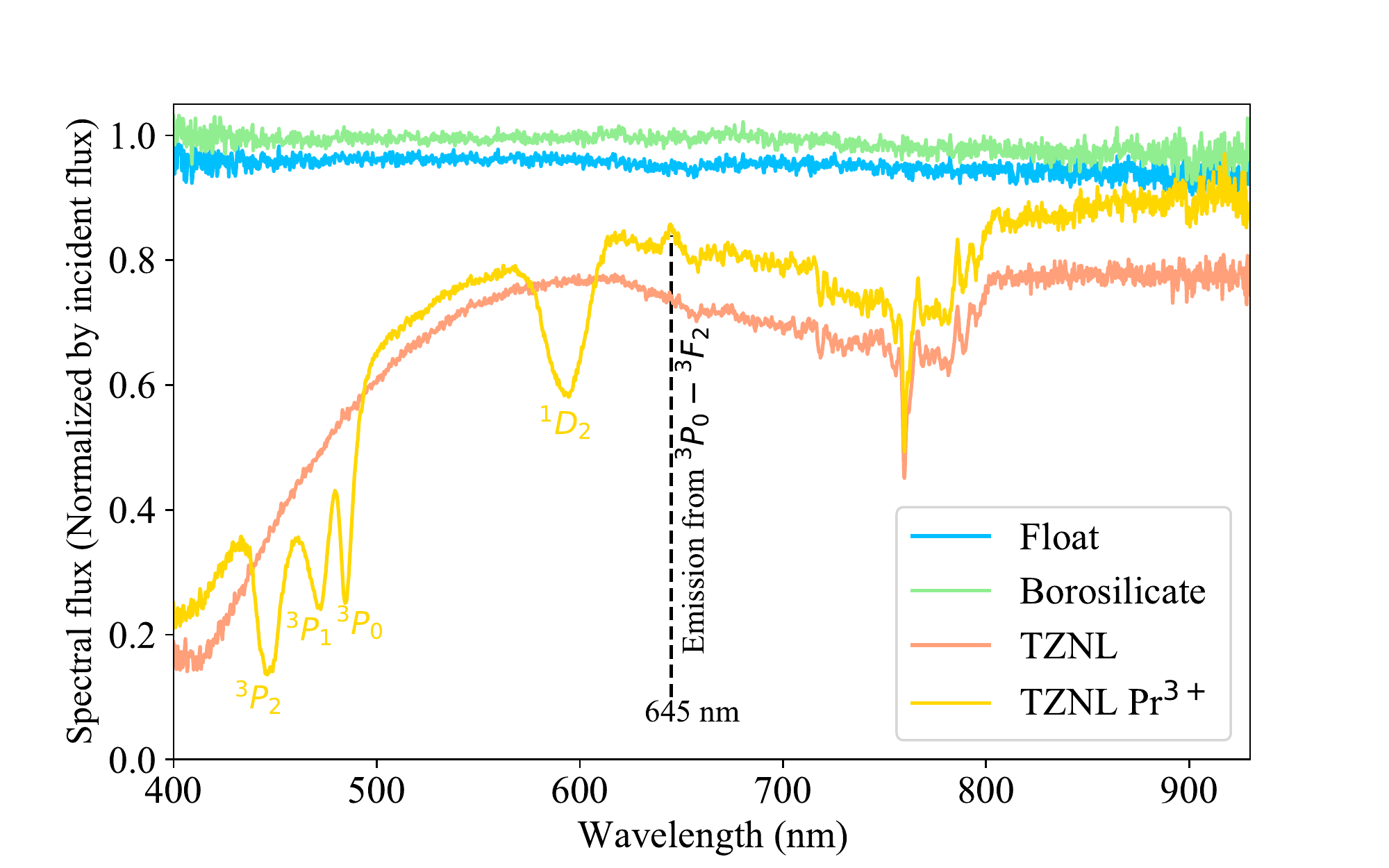}
\caption{Spectral flux under sunlight illumination for the samples investigated in this study.}
\label{fig:spec}
\end{figure}

A calibrated fiber-coupled spectrometer (LaserLine LSP-2) was used to measure the incident and transmitted spectral flux through the glass sample, under sunlight illumination. The results are shown in Fig.~\ref{fig:spec}, where the high quality of the thinner borosilicate sample is evidenced. In the Pr$^{3+}$ doped TZNL sample we could detect four ground state absorption bands from this dopant and an enhanced overall transmission of light. This may happen due to a small decrease in the refractive index and possibly due to reactions induced by the dopant~\cite{Taniguchi2020a}. A small band at 645 nm was also detected, and attributed to the $^3$P$_0$-$^3$F$_2$ decay of Pr$^{3+}$ ions. This result confirms that we have archived DF, still the ion is much more effective in absorbing than emitting photons.

\begin{figure}
\includegraphics[trim=1cm 1cm 1cm 1cm,scale=0.5]{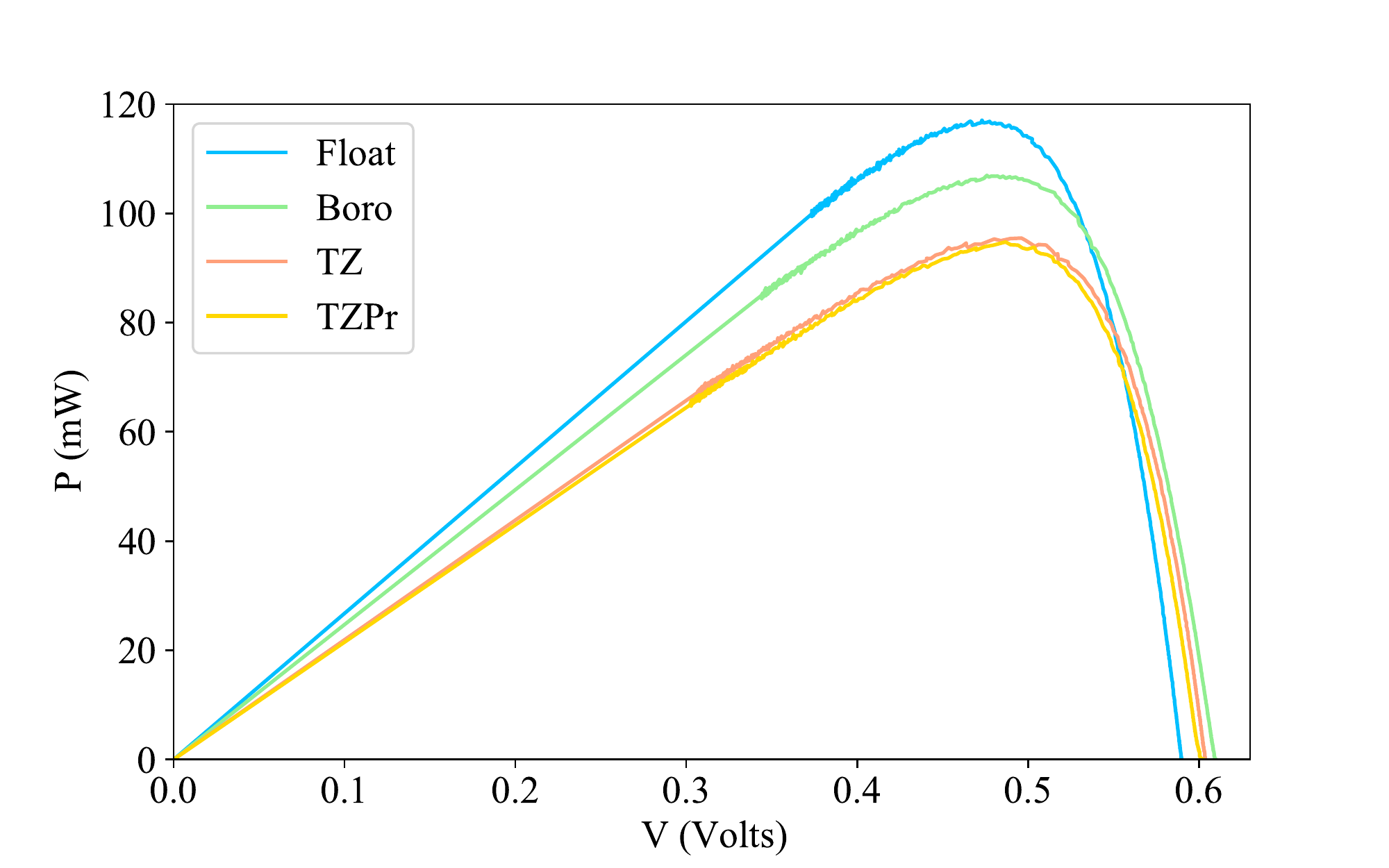}
\caption{Power of four solar cells encapsulated with different glasses}
\label{fig:pxv}
\end{figure}

The glasses were used to build prototypes. Four identical AR coated p-type Si solar cells (2.0x3.9 cm) were encapsulated at 85°C using ethylene-vinyl acetate (EVA) as the encapsulant and Tedlar-Pet-Tedlar (TPT) as the backsheet. The obtained prototypes were named Float, Boro, TZ, and TZPr, referring to the glass used in each one. A solar simulator (ScienceTech SS2.5kW) was used to measure IxV curves under standard conditions (AM1.5G spectrum, 1000 W/m$^2$ and cell temperature at 25°C), which were used to obtain the power as shown in Fig.~\ref{fig:pxv}.

The power achieved by TZ and TZPr was 95 mW, while Boro and Float have reached peaks of 106 mW and 116 mW, respectively. It is important to consider that the ASTM standard for solar simulators \cite{astm2019} is designed to compare silicon cells or modules under indoor testing. An AM1.5G spectrum in a class AAA simulator should reproduce the solar irradiance within a margin of $\pm$25\% in six bands of the spectrum (five 100 nm wide and the last one between 900-1100 nm). For optical spectroscopy, a solar simulator does not necessarily reproduce the solar spectrum. Also, other glass properties may affect the performance of the prototypes, and to properly evaluate them it is critical to measure their performances under real conditions, simultaneously. We developed a system to perform such measurements, which is described next.

From the data in Fig.~\ref{fig:pxv}, the circuit load to achieve peak power of $\sim$ 2 Ohms is obtained. Precision resistors (1 Ohm each) were used to obtain this load and Hall sensors (WCS2702) were used to measure the currents. A multimeter HP 34401a was used to obtain the sensitivity of the current sensors ($0.90\pm0.02$ mV/mA) and to measure the total load in each prototype ($2.08\pm0.05$ Ohm). Voltages were measured directly by the built-in analog-to-digital converter of a microcontroller (STM32103C8T6, resolution 0.8056 mV/level), which averaged current and voltage values over 192 measurements that took about 5 milliseconds each. The averaged values were sent to a computer via serial communication.

\begin{figure*}
\centering
\includegraphics[trim=0cm 0cm 0cm 0cm,scale=0.47]{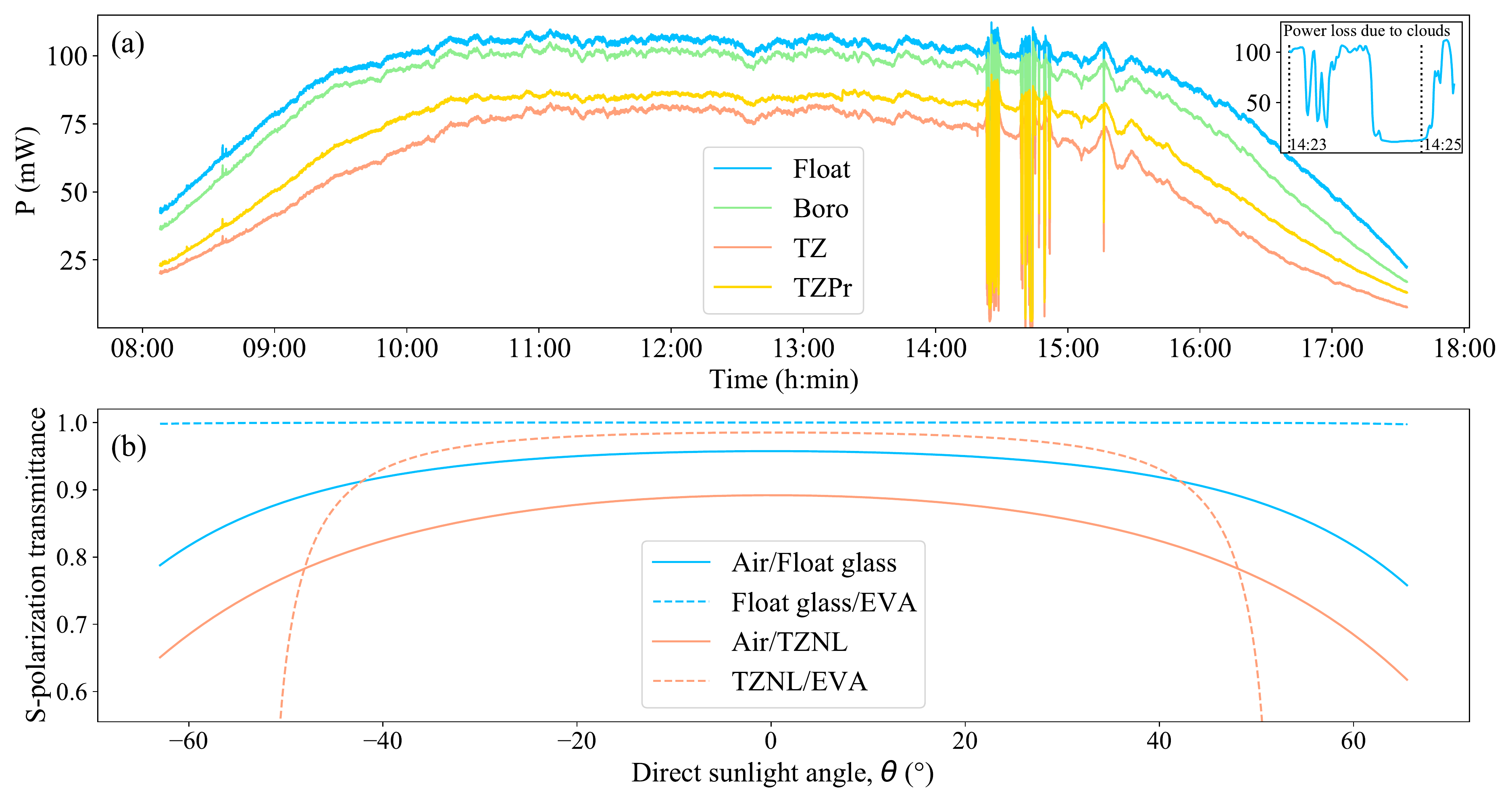}
\caption{(a) Power in each solar module prototype and (b) Fresnel transmittance of the s component (down) for several surfaces, as indicated. The time in (a) is in scale with the corresponding angle in (b).}
\label{fig:day}
\end{figure*}

The cells have to be cleaned and soldered to tinned copper wires before the encapsulation, where the components are heated at 85°C and slightly pressured. In the construction of the prototypes, all the steps may potentially introduce problems.  In order to evaluate our ability to reproduce all these steps, as well as the reliability of our measurement system, several prototypes were produced using float glass as cover. The power of four of these prototypes was simultaneously measured under sunlight illumination, and the results were consistent inside a margin error lower than 3\%. Damaged prototypes were intentionally evaluated and the power drop observed was very high if the silicon cell was cracked ($\sim$50-90\%), and measurable ($\sim$10\%) if air bubbles remained in the EVA below the cover glass. 
Figure~\ref{fig:day} (a) shows data obtained in a measurement performed on February 5 (2021, location -26.228°,-52.670°). The plane of the prototypes was set to be perpendicular to the sun around noon. The sunrise/sunset time was 06:09/19:21, which resulted in an angular change of about 0.2272°/min during 792 minutes on that day. The inset in this figure highlights the power loss due to clouds, which lowered momentarily the prototype temperatures. As a result, all devices reached their peak power on that day just after 14:25. To interpret the data,  the Fresnel transmittance of several surfaces in the prototypes was analyzed. Fig.~\ref{fig:day} (b) shows the S-polarization transmittance for all prototypes.

The high refractive index of the TZNL glass introduces a significant loss to the TZ and TZPr prototypes by the total internal reflection between TZNL and the EVA ($\theta_c\cong50^o$). Consequently, for direct sunlight angle higher than 50$^o$, TZ and TZPr prototypes allow only diffuse light to reach the silicon cell. Thus, under this condition, TZ and TZPr achieved only half of the power produced by the Float glass prototype. The Fresnel transmittance is shown in Fig.~\ref{fig:day} (b) highlighting the inflection point around the critical angle. In previous work, we have already discussed the challenges to use a high refractive index material as cover in a solar panel~\cite{Taniguchi2019}, and the results shown here confirm that. When the incident angle is low the performance of the prototypes remains dictated by the reflection losses, which are quite high in the TZNL matrix due to its high refractive index.

To achieve the total electrical energy produced by each device during the experiment, the power presented in Fig.~\ref{fig:day}~(a) was integrated for each prototype. A summary of the electrical output data obtained is shown in Table~\ref{tab:2}.
\begin{table}
    \caption{Peak power P(W) per area A(m$^2$) and energy produced E (J) during February 5 experiment, compared to the peak power in the solar simulator measurements (P$_s$) per area (A).}
    \centering
    \begin{tabular}{llll}\hline\hline
         & P/A (W/m$^2$) & E (J)  &P$_s$/A (W/m$^2$) \\
        \hline
        Float&143 (100\%)&3058 (100\%)&148 (100\%)\\
        Boro&137 (95\%)&2826 (92\%)&135 (86\%)\\
        TZ&110 (76\%)&2005 (65\%)&121 (82\%) \\
        TZPr&117 (82\%)&2288 (74\%)&121 (82\%) \\\hline\hline
    \end{tabular}
    \label{tab:2}
\end{table}
Float, boro, TZ and TZPr prototypes produced 3058, 2826, 2005, and 2288 Joules, respectively. As one can see, the TZ delivered only 65\% of the total power delivered by the Float, which is significantly lower than the 82\% observed under the solar simulator. This reduction in total energy is due to reflection losses in the TZNL/EVA interface, as we can see in Fig.~\ref{fig:day}~(b). While the s-polarization transmittance is near 1.0 for the Float/EVA interface during the entire day, transmittance at the TZNL/EVA interface is dramatically lower in the morning (before $\sim$10:00) and the afternoon (after $\sim$15:30). As a result, the power ratio between TZPr/Float increases from 53\% at the beginning of the measurement to 80\% around 10:00, which is just 2\% lower than what was achieved in the indoor testing. Such difference could be a result of different equilibrium temperatures in each module, once parameters such as thermal diffusivity and thickness should affect the heat dissipation, and consequently the silicon temperature and the performance of the devices. Such thermal investigation will be considered in future work.

The results shown here demonstrate the importance to investigate these cover glasses and the prototypes made with them under sunlight. The power measurement under the solar simulator is an important indicator of the prototype power in a real situation, though, as we have seen, the total energy produced may be significantly different. In our case, we could explain such differences mostly due to the s-polarization transmittance, as seen in Fig.~\ref{fig:day}, and even though the TZPr prototype performed better than the TZ, our results do not indicate that Pr$^{3+}$ and the DF related to it are playing an important role. On the other hand, the introduction of this dopant made the glass slightly more transparent, mainly in the infrared region where Si solar cells are more sensitive.

Glasses certainly may play a role to boost silicon solar cells. Nonetheless, if we aim to develop a practical solution to this challenge we cannot focus on spectral modifications only. For example, even if we could obtain a high quantum efficiency spectral converter based on a high refractive index material, such as the TZNL glass, it would be necessary to reduce reflection losses. This should require the use of a high refractive index polymer~\cite{Kleine2020}, for instance, which is likely to be unfeasible at the photovoltaic industry level. Besides that, several other requirements were not evaluated for the TZNL glass, such as its chemical or thermal strengthening, and the resulting mechanical resistance. TZNL is also an expensive glass, based in Tellurium oxide, which is a quite rare commodity~\cite{Bleiwas2010} and the mass production of this glass may not be feasible. However, in some specific applications, even expensive SC may be economically viable, as in concentrated photovoltaics~\cite{Wiesenfarth2018}, where sunlight is concentrated several hundred times in a small solar cell, and cover glass can be used.

Some authors have even discussed the rare-earth cost to produce a SC~\cite{Day2019}, but such values are meaningless if we are not accounting for the host and the process in which such SC would be produced and added to the silicon module. Allsopp et al.~\cite{Allsopp2020} have analyzed the performance of PV's prototypes made with several glass compositions quite similar to the float glass (doped versions). However, only solar simulator measurements were performed. As we have discussed, even a class AAA solar simulator does not exactly reproduce the sunlight spectrum, which could be important in order to evaluate a SC. As showed here, evaluating performances under real sunlight was possible using our homemade measurement system, and it could provide important and complementary information for PV characterization.

Pilkington describes in detail the challenges to develop the float process. In his words~\cite{Pilkington1969}: ``The principles of the process perhaps sound very simple,...How complex the development work was is illustrated by the fact that it took us seven years and four million pounds to make any saleable glass and a total of seven million pounds before the process could achieve the set objective of replacing plate''. Such an effort and industrial success of the float process were not achieved by chance, and after 50 years nothing more suitable to produce massive amounts of high-quality flat glass was discovered. 

In this context, the Pilkington method will continue to be fundamental for the tera-watt deployment of solar energy, still we could think of a modified glass to be produced by the same process. The method has already been used to fabricate other glasses, such as borosilicates~\cite{Wereszczak2014} and aluminosilicates~\cite{Li2017d,Zheng2017}, which are required in specific applications. Silicon solar cells are reaching a ``practical Schockley-Queisser limit''~\cite{McKenna2017a,Nayak2019}, and the projections made by industry~\cite{ITRPV2019} do not indicate any prospect that another technology will replace Si PVs. However, to achieve and sustain a multi tera-watt scale PV plant worldwide, the materials and processes involved should become more sustainable and cheaper. 

To the best of our knowledge, researchers have not yet explored the possibility of a low refractive index and low melting temperature glass. To produce 1 ton of float glass, about 5.5 GJ of energy is needed~\cite{Butler2011}, and a reduced melting temperature would result in energy savings as well as lower emissions of greenhouse gases.  If such material could achieve the chemical and mechanical resistance required for a cover glass in PVs, such devices would benefit from lower reflection losses in the air/glass interface that could be helpful at low incidence angles of light. SCs are an interesting technology, but at the industry level, we are restricted to use the float glass or the CVD process to obtain SCs. In future work, we will pursue different pathways to expand these possibilities.

\begin{acknowledgments}
The authors acknowledge the support from the Brazilian agencies CAPES, CNPq, Fundação Araucária and FINEP. Companhia Paranaense de Energia (COPEL) and Agência Nacional de Energia Elétrica (Aneel) are also acknowledged for the financial support through the contract P\&D 2866-0466/2017.
\end{acknowledgments}

\section*{Data Availability}
The data that support the findings of this study are available from the corresponding author upon reasonable request.

\section*{References}

\nocite{*}
\bibliography{aipsamp}

\end{document}